\documentclass[letterpaper, 10 pt, conference]{ieeeconf}  

\IEEEoverridecommandlockouts                              

\usepackage{graphicx}                                      
\usepackage{tikz}
\usepackage{pgfplots}
\usetikzlibrary{shapes,arrows,chains, automata, positioning}
\usepackage[caption=false,font=footnotesize]{subfig}       
\pgfplotsset{compat=1.18}
\usetikzlibrary{spy}

\usepackage{amsmath}                                       
\usepackage{amssymb}                                       
\usepackage{yhmath}                                        
\usepackage{mathtools}                                     
\usepackage{mdframed}                                      
\usepackage{array}                                         
\usepackage{multirow}

\usepackage[hidelinks]{hyperref}                           
\usepackage[capitalise]{cleveref}                          
\usepackage{soul}                                          
\usepackage{afterpage}
\usepackage{verbatim}                                      
\usepackage[normalem]{ulem}                                
\usepackage{float}          
\usepgfplotslibrary{units}
\usepackage[shortcuts,acronym]{glossaries}
\usepackage{flushend}

\newacronym{adass}{ADASs}{Advanced Driver Assistance Systems}
\newacronym{aebss}{AEBSs}{Autonomous Emergency Braking Systems}
\newacronym{aebs}{AEBS}{Autonomous Emergency Braking System}
\newacronym{av}{AV}{Autonomous Vehicle}
\newacronym{avs}{AVs}{Autonomous Vehicles}
\newacronym{lidar}{LiDAR}{Light Detection and Ranging Sensor}
\newacronym{ttc}{TTC}{Time-to-Collision}
\newacronym{more}{MORE}{Munich Mobility Research Campus}
\newacronym{nefimmore}{NeFiMMORE}{Near-Field Monitoring System MORE}
\newacronym{ros}{ROS}{Robot Operating System}
\newacronym{aabb}{AABB}{Axis-Aligned Bounding Box}
\newacronym{tpr}{TPR}{True Positive Rate}
\newacronym{fpr}{FPR}{False Positive Rate}
\newacronym{fnr}{FNR}{False Negative Rate}
\newacronym{fp}{FP}{False Positive}
\newacronym{fn}{FN}{False Negative}
\newacronym{cv}{CV}{Constant Velocity}
\newacronym{fov}{FOV}{Field-of-View}
\newacronym{ifd-vp}{IFD-VP}{Inter-Frame Displacement-based Velocity Predictor}

\DeclareGraphicsExtensions{.pdf, .png}                      

\title{\LARGE \bf
Improving Functional Reliability of Near-Field Monitoring for Emergency Braking in Autonomous Vehicles
}

\author{Junnan Pan$^{1}$, Prodromos Sotiriadis$^{1}$, Vladislav Nenchev$^{1}$, Ferdinand Englberger$^{1}$, \textit{Member, IEEE}
\thanks{$^{1}$Junnan Pan, Prodromos Sotiriadis, Vladislav Nenchev, Ferdinand Englberger are with Department of Electrical and Computer Engineering, University of the Bundeswehr Munich, Germany (Corresponding author: Junnan Pan),
        {\tt\small \{junnan.pan, sotiriadis.prodomos, vladislav.nenchev, ferdinand.englberger\}@unibw.de }}
}

\begin{document}

\setulcolor{red}

\pgfdeclarelayer{marx}
\pgfsetlayers{main,marx}
\providecommand{\cmark}[2][]{%
  \begin{pgfonlayer}{marx}
    \node [nmark] at (c#2#1) {#2};
  \end{pgfonlayer}{marx}
  } 
\providecommand{\cmark}[2][]{\relax} 

\maketitle
\thispagestyle{empty}
\pagestyle{empty}

\begin{abstract}
Autonomous vehicles require reliable hazard detection. However, primary sensor systems may miss near-field obstacles, resulting in safety risks. Although a dedicated fast-reacting near-field monitoring system can mitigate this, it typically suffers from false positives. To mitigate these, in this paper, we introduce three monitoring strategies based on dynamic spatial properties, relevant object sizes, and motion-aware prediction. In experiments in a validated simulation, we compare the initial monitoring strategy against the proposed improvements. The results demonstrate that the proposed strategies can significantly improve the reliability of near-field monitoring systems.
\end{abstract}

\section{Introduction} \label{sec:introduction}
\ac{adass} are rapidly advancing toward improved automation and safety in transportation; however, ensuring robust and reliable safety of self-driving vehicles remains a critical challenge. While \ac{aebss} have demonstrated potential in reducing collision risks, these systems often rely on the same sensor setup as the high-level driving system, which can miss hazards outside the sensors' \ac{fov}. Especially in complex urban environments, such blind spots underscore the need for dedicated near-field monitoring systems as an additional safety layer. 

In previous work, a microcontroller-based near-field monitoring system with a dedicated sensor was tailored to cover the blind spots of a primary sensor suite of an \ac{av} \cite{junnan_pan_collision_2023, pan_independent_2024}. While effective in its fast reaction capability, this system exhibited oversensitivity, resulting in frequent \ac{fp} braking events that could disrupt normal driving and decrease user confidence. The challenges originate from constraints of the embedded solution, where computational resources and sensor resolution are limited.

To address these challenges, this paper proposes a set of advanced monitoring strategies to mitigate \ac{fp} braking events. Key improvements include establishing a spatial corridor limited to the sensor’s \ac{fov} and a speed-dependent monitoring boundary that incorporates both the reaction and braking distances to filter out irrelevant data. Further, an object-size threshold is used to distinguish between hazardous obstacles and negligible objects (e.g., small debris or insects). Finally, motion prediction methods (with a focus on low-complexity approaches suitable for microcontroller platforms) are used to maintain awareness of moving objects in an anticipating manner. The strategies are tested using a validated simulation environment that has been quantitatively compared with real-world sensor data, ensuring that the simulated conditions are representative of practical scenarios. The experimental results demonstrate that these improvements not only reduce \ac{fp} triggers significantly but also maintain the critical responsiveness required for emergency braking applications. 

The contributions of this work are threefold:
\begin{itemize}
\item We propose three computationally efficient monitoring strategies;

\item We present an evaluation wherein simulated sensor data is validated against laboratory measurements;

\item We demonstrate a significant functional reliability improvement for near-field monitoring in \ac{av}s.
\end{itemize}

The remainder of this paper is organized as follows.  \cref{sec:literature} provides related work followed by the problem statement (\cref{sec:problem_statement}). \cref{sec:monitoring_strategies} presents the monitoring strategies. \cref{sec:eval_methodology} presents the simulation setup and evaluation results, followed by a conclusion (\cref{sec:conclusion}).

\section{Related work}\label{sec:literature}
An \ac{aebs} perceives the traffic environment ahead by leveraging onboard sensors, such as millimeter-wave radar, cameras, and \ac{lidar}. A cascaded braking process is applied by evaluating impact factors, such as the \ac{ttc}. Collisions can be avoided through braking actions, ranging from partial to full braking. In the work of \cite{alsuwian_autonomous_2022}, a data fusion method is applied to the \ac{aebs} to improve the reliability of sensor data, which can also prevent \ac{fp}s caused by single-sensor errors. However, as the authors addressed in their work \cite{yang_systematic_2022}, the current \ac{aebs} may be less effective or optimized in situations where a side-moving object suddenly intersects with the ego vehicle's driving lane or during vehicle turning, as the \ac{aebs} is primarily designed for front-moving targets in the longitudinal direction. Various object detection and segmentation techniques have also been explored to enhance the fidelity of environmental monitoring. 

Clustering methods have become central to segmenting \ac{lidar} point clouds into meaningful object representations. Two widely used approaches are Euclidean distance clustering and density-based clustering  \cite{reich_low_2024, zhang_vehicle_2020}. The Euclidean clustering is, due to its low complexity and simple implementation, suitable for embedded applications.

For passengers, it is more convenient to avoid imminent collisions through smooth steering actions rather than sudden braking. Recent research investigates collision avoidance by utilizing set-based reachability \cite{pek_fail-safe_2021} or analytical safety bounds \cite{jacumet_analytical_2023}. A collision-free evasive trajectory can be generated by exploring the reachability set of the automated vehicle in the forward driving area. The reachability set of the automated vehicle comprises all possible reachable states from an initial set of states \cite{xausa_verification_2015}. 

Active safety systems in \ac{av} also implement object monitoring filtering techniques, utilizing attributes such as space, size, and motion to differentiate between genuine collision hazards and innocuous objects or sensor noise. \cite{yatbaz_integrity_2024} constrained the input processing lidar point cloud for the deep neural networks by employing spatial filtering for 3D object detection. It is evident that this results in the system's focalization on the proximate objects of the ego vehicle.



\section{Problem Statement} \label{sec:problem_statement}



The drive towards autonomy in \ac{adass} raises safety requirements for increasingly challenging driving environments. Traditional collision avoidance systems (such as \ac{aebs}) primarily focus on obstacles detected within their primary sensor \ac{fov}. However, these sensors may leave critical safety gaps due to limited primary sensor coverage. A near-field monitoring system can provide an additional safety layer by detecting obstacles with dedicated hardware, as described in the following. The necessity for additional safety measures is becoming increasingly apparent as the end-to-end \ac{av} is frequently interpreted as a “black box”, thereby complicating safety assurance.

\subsection{System Description} \label{sec:system_description}
To enhance the safety of an \ac{av}, a near-field monitoring system was developed for the \ac{more}  project \cite{junnan_pan_collision_2023, pan_independent_2024}. The \ac{nefimmore} is designed to provide an additional safety layer for autonomous driving through the use of a dedicated 3D \ac{lidar}. The goal is to develop an active safety system with a low-cost embedded solution that could eventually be integrated directly into the \ac{lidar} sensor.

An autonomous driving system generates a collision-free trajectory to drive the vehicle safely along this path to its destination. In some scenarios, the autonomous driving system may not respond quickly enough to avoid collisions due to unexpected events, such as a child suddenly crossing from an occluded sensor detection area. The independent \ac{nefimmore} system is designed to handle such scenarios by applying emergency braking to avoid collisions or reduce their severity if unavoidable. A low-positioned 3D \ac{lidar} at the front of the vehicle is used to perceive the near-field environment. The front \ac{lidar} can detect blind spots of the high-positioned \ac{lidar} of the primary sensor system.
\begin{figure}[t!]
  \centering
  \subfloat[Ego vehicle overtaken.\label{fig:overtaking}]{
    \begin{minipage}[b]{0.45\columnwidth}
        \centering
        \includegraphics[width=0.53\columnwidth]{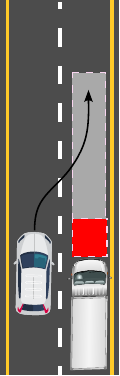}
    \end{minipage}
  }
  \hfill
  \subfloat[Ego vehicle driving in the curve.\label{fig:curve_monitoring}]{
    \begin{minipage}[b]{0.45\columnwidth}
        \centering
        \includegraphics[width=1\columnwidth]{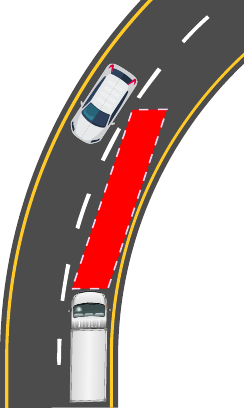}
    \end{minipage}
  }
  \caption{The keep-out area (marked in red) for monitoring in typical driving scenarios.}
  \label{fig:monitor_scenario}
\end{figure}

The system's monitoring of the front driving area is illustrated for two typical driving scenarios in \cref{fig:monitor_scenario}. The keep-out area is marked in red. It varies along the driving corridor, primarily depending on the current vehicle speed, as speed directly affects the braking distance of the ego vehicle.

\subsection{Problem Formulation}
Despite its rapid response time, the initial implementation suffered from a high \ac{fpr} due to its oversensitivity. In particular, the simple single-point hit detection method triggered unnecessary emergency braking, thereby affecting normal vehicle operation and passenger comfort. The core problems we have to tackle are to reduce \ac{fp} braking triggers, to maintain limited computational resources (since the system is designed for microcontroller-based implementations), to distinguish between static and dynamic hazards, and to maintain real-time responsiveness under uncertainty. 

\section{Advanced Monitoring Strategies} \label{sec:monitoring_strategies}
This section proposes three integrated monitoring strategies to reduce \ac{fp}s while maintaining a rapid hazard response. These strategies have been developed to align with our embedded platform, which strikes a balance between computational efficiency and robust detection capabilities. Since the focus of this work is to verify the proposed monitoring strategies, and as lateral constraints in curved corridors continuously change, increasing the geometric complexity and complicating the verification of the methods, we focus our analysis on straight-line driving scenarios in low-speed urban environments.

\subsection{Dynamic Spatial Range Adaptation} \label{subsec:spatial_constraint}

To ensure timely and reliable reactions, the monitoring system must continuously adapt its monitoring range based on the vehicle’s speed. The sum of the reaction distance $D_r$ and braking distance $D_b$ defines the vehicle's total stopping distance. If an object appears within this stopping distance area, a collision with the object may not be avoided without a steering maneuver. To introduce basic collision avoidance capability, we extend the monitoring distance by adding a safety margin distance $D_s$. Thus, the monitoring range of the system can be calculated using ~\cref{eq:monitor_distance}. All components contributing to the monitoring distance are velocity-dependent. Specifically, the reaction distance and the safety margin distance are both proportional to the vehicle velocity $v$. The system reaction time $t_r$ represents the decision time of the system. The safety margin time $t_s$ adds a small buffer period to help avoid potential collisions in advance.

\begin{align}
    D_{Mon}(v) & = D_r(v) + D_s(v) + D_b(v) \label{eq:monitor_distance}\\
               & = (t_r + t_s)v + D_b(v). \notag
\end{align}

To calculate the braking distance, we assume that the maximum torque is applied to the tire and use \cref{eq:braking_distance}. In this study, a typical dry asphalt friction coefficient of $\mu = 0.75$ is adopted for evaluation. The braking distance formula is derived from fundamental physical principles of deceleration \cite{poul_greibe_braking_2007} and incorporates both the friction coefficient and the effect of road slope. In the case of a vehicle traveling on a perfectly horizontal surface, the road slope $s$ is zero.
\begin{equation}
    D_b(v) = \frac{v^2}{2 g (\mu + s)}.
    \label{eq:braking_distance}
\end{equation}
where $g$ stands for the acceleration due to gravity.

One of the objectives of the \ac{nefimmore} active safety system is to ensure a collision-free driving path, which is generated by the path planner. Therefore, except for the restriction of the frontend boundary, we also restrict the lateral and vertical boundaries of the front driving area. The lateral and vertical boundaries build a spatial driving corridor in which the autonomous vehicle should safely move forward without collision. Outside the corridor, there is less interest in the system. This additional monitoring constraint can therefore reduce unnecessary perception and monitoring of the environment. Combined with dynamic longitudinal monitoring, the lateral and vertical constraints significantly reduce unnecessary computation by limiting the region of interest.

Due to the geometric simplicity of a straight road, the lateral boundary $y_{lat}$ can be expressed as constant values using the following boundary expression:
\begin{equation}
    y_{lat}=
    \begin{cases}
       \frac{w}{2}, &\text{left boundary}\\
      -\frac{w}{2}, &\text{right boundary}
     \end{cases}
\end{equation}
where $w$ denotes the outermost width of the vehicle. At higher vehicle speeds, the lateral boundary can be extended to allow for slight deviations from the straight path, providing fault tolerance. However, in our low-speed urban application scenario, this effect can be considered negligible.

The vertical boundaries $z_{ver}$ are determined by the geometry of the vehicle, and it can be defined as constant height values. Specifically, they are given by:
\begin{equation}
    z_{ver} =
    \begin{cases}
       h_{chassis}, &\text{bottom boundary}\\
       h_{top}, &\text{top boundary}
     \end{cases}
\end{equation}
where $h_{chassis}$ denotes the height of the vehicle chassis and $h_{top}$ represents the topmost height of the vehicle. Introducing the bottom boundary helps exclude ground-level laser points when the vehicle is traveling on a flat surface. Thus, the common ground removing algorithm step in the pre-processing of point clouds can be skipped under certain circumstances. Similarly, the top boundary constrains the field of view above the vehicle, reducing the inclusion of unnecessary data.

By limiting the \ac{lidar}'s \ac{fov}, the monitoring region of interest is reduced to the driving corridor. As a result, the number of point cloud points can be significantly reduced, which lowers the computational load for data processing. The size of the corridor is dependent on the geometry of the vehicle. For larger vehicle types, such as trucks or transporters, the corridor is correspondingly wider.

\subsection{Size-based Object Filtering} \label{subsec:strategy:size-based}
The single-point hit method for obstacle detection, which is used in the previous work \cite{pan_independent_2024}, due to its slack policies — reaction whenever any single \ac{lidar} point enters the monitoring zone -- shows its significant advantages in the fast response. However, the biggest drawback of this method is the maximum sensitivity towards any size object in the monitoring area. In practice, it is neither acceptable nor comfortable that the autonomous vehicle applies emergency braking multiple times without any true hazards in the driving direction.

Utilizing the geometric constraints previously delineated, a size-based object filter (or balloon filter) is employed to mitigate sensitivity to obstacles. Within the sensor monitoring sight range defined by the preceding fundamental methods, an object size restriction could be specified. Any objects within this restriction of size should be ignored. In this way, the \ac{fp}s due to small debris, flying insects or swirling dust could be avoided. The single-point hit method is a special case of the size-based filter, which has a minimum size for the obstacle.

The purpose of the size-based filter is to reduce the influence of the small objects on the collision check. The basic requirements for selection of the filter size are that the traffic users in the urban driving environment, like pedestrians and cars, should not be excluded (too large size may increase the \ac{fnr}), and insignificant small objects like fallen leaves and footballs should be excluded (too small size may increase the \ac{fpr}). From these aspects, we choose an appropriate filter size of 0.3 m. A cluster is discarded only if all three dimensions -- height $h$, width $w$ and depth $d$ -- are smaller than 0.3 m. Within this size, the small flying animals, balls, and most of the fallen tree branches can be excluded from the collision check. Above this size, like dog-sized/cat-sized animals and all the traffic users can be effectively left. Correspondingly, an object is kept whenever at least one of its extents satisfies $\max(h,w,d) \geq 0.3\ \text{m}$. 

\begin{figure} [!t]
    \centering
    \includegraphics[width=1\columnwidth]{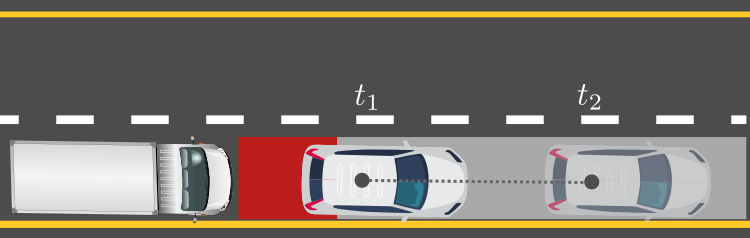}
    \caption{Schematic of motion-aware predictive monitoring. The target vehicle moves from time $t_1$ to time $t_2$.}
    \label{fig:schemmatic_pre-monitoring}
\end{figure}

\subsection{Motion-aware Predictive Monitoring}
Using the strategies presented in \cref{subsec:strategy:size-based} irrelevant small objects as well as some spurious detections can be ignored. However, these filtering approaches only consider the spatial constraint of the object and do not account for object movement. In a car-following scenario, the distance between the ego vehicle and the object ahead can sometimes fall within the monitored driving corridor. If the ego vehicle's velocity remains equal to or lower than the front target, the collision probability is low and there should be no need to trigger emergency braking. 

To tackle the aforementioned limitations, we propose an additional monitoring strategy that accounts for object motion by extending the monitoring range. The schematic of this monitoring is illustrated in \cref{fig:schemmatic_pre-monitoring}. For the size-based filter detection method, the ego vehicle should take action at the time $t_1$. However, as the front target vehicle keeps away from the ego vehicle at the time $t_2$, the \ac{nefimmore} system should retain silence in such case. The extended monitoring area provides additional reaction space, allowing the system to predict the objects based on the object motion. 

In the car-following scenario, the relative velocity between the front target and the ego vehicle in the longitudinal direction is calculated and utilized as a potential indicator for collision. If the relative velocity is larger or equal to zero, the likelihood of the potential collision can be negligible, even when the relative distance falls in the braking distance. Without considering the motion properties of the object, the spatially limited strategies can trigger an unnecessary braking signal in such cases. Conversely, suppose the relative velocity is less than zero, but the relative distance is within the braking distance. In that case, a potential collision will be detected and the emergency braking will be applied.

\section{Evaluation} \label{sec:eval_methodology}
This section outlines a validation framework that is used to evaluate the reliability of the \ac{nefimmore} system. It includes the design and implementation of specialized simulation platforms for experimenting with the aforementioned strategies under controlled driving scenarios. Subsequently, targeted, designed scenarios are constructed to methodically assess the proposed enhancement monitoring strategies.

The main purpose of the strategy in the \cref{subsec:spatial_constraint} is to establish fundamental spatial geometric constraints before the implementation of other advanced monitoring strategies. They are used to intuitively reduce irrelevant objects from the monitoring \ac{fov} and to further reduce the computational overload associated with the deployment of embedded systems. Therefore, separate simulation experiments for this category are not conducted. Instead, their role is inherently validated and implicitly captured within the more comprehensive evaluations of advanced false-positive reduction strategies, which inherently depend on these basic geometric constraints.

\subsection{Simulation Validation}
The primary objective of the simulation validation at the beginning of this evaluation is to establish the trustworthiness of the simulated \ac{lidar} data. By quantitatively comparing the simulated 3D-\ac{lidar} point cloud environment with real-world data collected in the laboratory, we aim to demonstrate that the simulation environment has sufficient accuracy for this evaluation process. The validation of this approach is pivotal for the subsequent utilization of simulations. Such simulations are essential for rigorously evaluating and verifying the effectiveness of our proposed monitoring strategies. This is achieved under controlled and reproducible conditions.

In both the laboratory and simulation settings, the \ac{lidar} sensor (Velodyne VLP-16) is mounted in an inverted orientation, matching the actual vehicle installation. A dedicated front space in the laboratory (extending up to 8 m in front and 1 m on each side) of the sensor is defined as the region of interest. %
The virtual environment used for simulating object and sensor behavior in this evaluation is the Gazebo robotics simulator. The \ac{ros} provides the communication interface between the Gazebo simulation environment and the algorithms implementing the proposed strategies. The VLP-16 in the simulation is configured to have 1875 horizontal samples from -180\textdegree{} to 180\textdegree{} with approximately 0.2\textdegree{} resolution, 16 vertical samples from -15\textdegree{} to 15\textdegree{} with 2\textdegree{} resolution, a sensor update rate of 10 Hz, and a Gaussian noise standard deviation of 0.008. Considering the controlled laboratory environment, we did not introduce artificial environmental noise in the simulation. Since our primary aim is to verify the business logic of these strategies, we simplify the ego vehicle as a point model, placing the \ac{lidar} sensor at this point.

\begin{figure} [t]
    \centering
    \begin{tikzpicture}
    \begin{axis}[
        name=main,  
        width=0.9\columnwidth,
        height=0.5\columnwidth,
        grid=both,
        xlabel={Distance [cm]},
        ylabel={Number of Hits},
        legend style={at={(0.5,-0.4)}, anchor=north, legend columns=-1},
        xtick distance=120,
        ymajorgrids=true,
        grid style=dashed,
        enlargelimits=0.05,
        tick label style={font=\small},
        label style={font=\small},
        legend cell align={left},
    ]
    
    \addplot[
        color=blue,
        mark=*,
        thick,
    ]
    coordinates {
        (60,2033)(120,570)(180,250)(240,130)(300,104)(360,62)(420,36)
        (480,32)(540,28)(600,25)(660,24)(720,21)(780,20)
    };
    \addlegendentry{Real Hits}
    
    \addplot[
        color=red,
        mark=square*,
        thick,
        dashed,
    ]
    coordinates {
        (60,2032)(120,520)(180,258)(240,132)(300,100)(360,42)(420,38)
        (480,34)(540,30)(600,26)(660,22)(720,22)(780,18)
    };
    \addlegendentry{Simulated Hits}
    
    \node[draw, rectangle, black, thin, minimum width=10pt, minimum height=10pt] (rect120) at (axis cs:120,545) {};
    \node[draw, rectangle, black, thin, minimum width=10pt, minimum height=8pt] (rect360) at (axis cs:360,52) {};
    
    \end{axis}
    
    \begin{axis}[
        name=inset120,
        at={(main.north)},
        anchor=north west,
        xshift=10pt,
        yshift=-10pt,
        width=0.3\columnwidth,
        height=0.25\columnwidth,
        xlabel={},
        ylabel={},
        xmin=90, xmax=150,
        ymin=500, ymax=600,
        xtick={120},
        ytick={520,570},
        tick label style={font=\tiny},
        grid=both,
        grid style=dotted,
    ]
    
    \addplot[
        color=blue,
        mark=*,
        thick,
    ]
    coordinates {
        (120,570)
    };
    
    \addplot[
        color=red,
        mark=square*,
        thick,
        dashed,
    ]
    coordinates {
        (120,520)
    };
    \end{axis}
    
    \begin{axis}[
        name=inset360,
        at={(main.north east)},
        anchor=north east,
        xshift=-10pt,
        yshift=-10pt,
        width=0.3\columnwidth,
        height=0.25\columnwidth,
        xlabel={},
        ylabel={},
        xmin=330, xmax=390,
        ymin=30, ymax=70,
        xtick={360},
        ytick={42,62},
        tick label style={font=\tiny},
        grid=both,
        grid style=dotted,
    ]
    
    \addplot[
        color=blue,
        mark=*,
        thick,
    ]
    coordinates {
        (360,62)
    };
    
    \addplot[
        color=red,
        mark=square*,
        thick,
        dashed,
    ]
    coordinates {
        (360,42)
    };
    \end{axis}
    
    \draw[->, thin, black] (inset120.south west) -- (rect120.north east);
    \draw[->, thin, black] (inset360.south west) -- (rect360.north east);
\end{tikzpicture}
    \caption{Comparison of the static validation between real-world data and simulation data.}
    \label{fig:sim_valid:baseline_static_valid}
\end{figure}
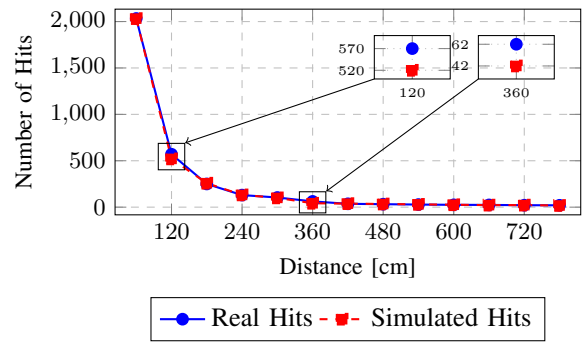

We designed a static validation procedure to assess the reliability of the simulation. In the static validation, we place a standard cuboid with known dimensions of 26 cm (width) $\times$ 36 cm (height) $\times$ 12 cm (depth) above the ground in the middle front of the sensor, supported by a platform at a certain height. This supported platform will be invisible to the system due to the  previously mentioned restricted bottom boundary (as described in \cref{subsec:spatial_constraint}). This cuboid is positioned at various distances from 1.2 m to 7.8 m, with a step of 0.6 m for each trial. For each measurement, we count the point number of the cuboid and compare with the value in the corresponding simulation measurement.

The result of the point detected hits comparison of scenarios for this static validation is illustrated in \cref{fig:sim_valid:baseline_static_valid}. The detected point number in the region of interest observed on the surface of the cuboid box in the real-world measurement is closely matching the result in the simulation for each experiment trial. In addition to the comparison of the point hits on the object presented in the diagram, by comparing the point distribution (not shown in this paper), the real-world measurement and simulation also show the similarity of the point distribution on the cuboid box. 

This close consistency under static conditions confirms that the simulated sensor behavior can replicate the physical reality. Furthermore, no systematic result bias of point density in the simulation is observed, suggesting that the simulation environment is accurate enough to be used as a trustworthy baseline for evaluating the object for further experiments.

\subsection{Size-based Filter Evaluation} \label{subsec:eval:size-based}
In the evaluation of the size-based filtering strategy, the motion of the ego vehicle does not influence the accuracy of the results. Therefore, the ego vehicle remains stationary throughout the simulation. Instead, objects of varying sizes are introduced into the monitoring corridor at different distances relative to the ego vehicle's position. This setup isolates the effect of object size on detection performance and ensures repeatability across test scenarios.

Since the focus of this paper is evaluating the filtering strategy rather than optimizing clustering performance, we employ a standard Euclidean-based clustering algorithm to segment the point cloud into individual clusters. During clustering, if the \ac{aabb} of a forming cluster exceeds a predefined maximum dimension $D_{\text{max}}$, the cluster is flagged as a potential hazard. Although using an \ac{aabb} can lead to conservative size estimates (due to alignment with global axes and loss of object orientation information), it provides a computationally efficient approximation suitable for our system. 

The dimensions of the cube target set used in this simulation experiment are as follows: \{5, 10, 15, 20, 25, 30, 35, 40, 45, 50, 55, 60\}\,cm. Each target was traversed horizontally from outside the monitoring area through the monitoring area at the initial position, ranging from 1.2 m to 7.8 m with a 0.6 m distance step from the front \ac{lidar} of the ego vehicle. There were a total of 144 trials. The cube was positioned at a height of 0.6 m from the ground. We utilized the dataset collected during these experimental trials to assess the effectiveness of the proposed method. As previously mentioned in~\cref{subsec:strategy:size-based}, the system employs a threshold of 30 cm as the minimum size required for an object to be categorized as an obstacle. Objects below this dimension are considered non-hazardous. 

The result of the simulation evaluation in \cref{tab:size-based_filter:compare_result} indicates that the slack single-point hit detection strategy has a high \ac{fpr} of 70\% and high \ac{tpr} of 100\%, particularly for small targets, but maintains a F1-score of 66.67\%. The F1-score is the harmonic mean of precision and recall, and it provides a balanced measure of accuracy. In contrast, the proposed size-based object filter strategy reduces the FPR to 0\% under this specific experimental setup, while keeping the F1-score at 100\%. With the proposed size-based object filter, we can avoid this unwanted braking action caused by small objects less than 30 cm. It should be noted that this filtering approach may introduce false negative events. This occurrence arises due to the sparse property of the point cloud when the object is located at greater distances. The limited number of points cannot comprehensively represent the object geometry at far distances. For near distances, as is the case in the application of our system, this effect may be minimal, and even when the target is not detected due to the further distance, it will still be detected as the ego vehicle approaches.
\begin{table}[t]
    \centering
    \caption{Comparison of \ac{fpr}, \ac{tpr}, F1-score between the single-point hit strategy and Size-based object filter strategy.}
    \label{tab:size-based_filter:compare_result}
    \resizebox{\columnwidth}{!}{%
        \begin{tabular}{|l||cc|}
        \hline
         Measures  & Single-point Hit  & Size-based Object Filter \\ \hline
        \ac{fpr}   & 70\%       & 0\%                  \\ 
        \ac{tpr}   & 100\%      & 100\%                \\ 
        F1-score   & 66.67\%    & 100\%                \\ \hline
        \end{tabular}%
    }
\end{table}

\subsection{Motion-aware Monitoring Evaluation}
In the motion-aware evaluation, we assess the effectiveness of incorporating the motion-awareness strategy in the Gazebo simulation. The clustering algorithm used to generate bounding boxes, previously described in \cref{subsec:strategy:size-based}, is also applied here. Since this strategy considers object motion, clusters from consecutive \ac{lidar} frames must be matched to determine whether the same object is moving within the monitoring area. We use the Hungarian algorithm \cite{wang_tracking-by-detection_2022} for this matching step, as it efficiently aligns clustered objects between frames within the constrained driving corridor. To reduce computational complexity, we also apply a physical motion constraint assumption: objects (moving or stationary) must remain within a realistic positional range between consecutive frames, as determined by the allowed maximum velocity. Consequently, clustered objects falling outside this range are excluded from the matching algorithm.

We designed two different constant-speed car-following scenarios in the simulation: Constant-Gap following and Increasing-Gap following. Since the focus of this paper is to analyze the effect of the newly proposed methods on \ac{fp}s, the Decreasing-Gap following scenario (negative relative velocity), which would eventually lead to a collision, is outside the scope of this paper. For the cases of our experiments, there are no true collision events. Therefore, we will also not conduct the comparison of \ac{tpr} and \ac{fnr} for this approach. In the Constant-Gap following scenario, the relative velocity between the ego vehicle and the object is zero, meaning the relative distance remains nearly constant. In this scenario, the system should maintain awareness without triggering emergency braking. In the Increasing-Gap following scenario, the relative velocity of the ego vehicle is positive, indicating that the gap between the ego vehicle and the object is increasing. In this scenario, the system must evaluate whether emergency braking is necessary. In the Constant-Gap scenario, both the lead vehicle and ego vehicle maintain the same constant speed, ranging from 5 km/h to 30 km/h in increments of 5 km/h. In the Increasing-Gap scenario, the lead vehicle's speed ranges from 10 km/h to 30 km/h in increments of 5 km/h, while the ego vehicle's speed is consistently 5 km/h slower than the lead vehicle's. For each scenario, the initial gap between the lead vehicle and the ego vehicle varies from 2 m to 10 m in increments of 2 m. For each scenario, there were 30 trials. In this evaluation, for the controlled and simplified driving scenarios, the motion of the front object is predicted by analyzing the inter-frame displacement.

The comparison of \ac{fpr} between the size-based filtering strategy and the motion-aware strategy is presented in \cref{tab:motion-aware:compare_result}. In the baseline size-based filter, it triggers \ac{fp}s in almost half of the Constant-Gap runs. The results indicate that the motion-awareness predictive monitoring strategy can effectively reduce the \ac{fpr} for each car-following scenario. Although the motion-aware method may introduce extra computation cost, it improves the reliability of the near-field monitoring to avoid unnecessary triggers.
\begin{table}[t]
    \centering
    \caption{Comparison of \ac{fpr} between the size-based object filter strategy and motion-aware strategy.}
    \label{tab:motion-aware:compare_result}
    \resizebox{\columnwidth}{!}{%
        \begin{tabular}{|l||cc|}
        \hline
             Scenario     & Size-based filter & Motion-aware  \\ \hline
         Constant-Gap     & 46.7 \%           & 4.2 \%        \\ 
        Increasing-Gap    & 28 \%             & 0 \%          \\ \hline
        \end{tabular}%
    }
\end{table}

\section{Conclusions} 
\label{sec:conclusion}

We have shown that combining dynamic spatial range adaptation, size-based filtering, and motion-aware predictive monitoring improves reliability in near-field emergency braking. Through validated simulations against real-world sensor data, our microcontroller-compatible methods demonstrated a substantial reduction in \ac{fp}s, while preserving true-positive detection. Limitations remain under sparse point-cloud conditions (e.g., heavy rain) and when processing highly irregular obstacles; we plan to integrate adaptive noise modeling and a Kalman filter to address these. Future work will extend validation to curved roads and adverse weather in real-vehicle trials and perform formal safety verification, e.g., using \cite{nenchev_compositional_2025}, to further enhance system safety.



\addtolength{\textheight}{-12cm}   




\section*{ACKNOWLEDGMENT}
This research is funded by dtec.bw – Digitalization and Technology Research Center of the Bundeswehr, which we gratefully acknowledge.
dtec.bw is funded by the European Union -- NextGenerationEU.

\bibliographystyle{IEEEtran} 

\bibliography{Nahfeldueberwachungssystem-dtec-more}

\begin{thebibliography}{10}
\providecommand{\url}[1]{#1}
\csname url@samestyle\endcsname
\providecommand{\newblock}{\relax}
\providecommand{\bibinfo}[2]{#2}
\providecommand{\BIBentrySTDinterwordspacing}{\spaceskip=0pt\relax}
\providecommand{\BIBentryALTinterwordstretchfactor}{4}
\providecommand{\BIBentryALTinterwordspacing}{\spaceskip=\fontdimen2\font plus
\BIBentryALTinterwordstretchfactor\fontdimen3\font minus \fontdimen4\font\relax}
\providecommand{\BIBforeignlanguage}[2]{{%
\expandafter\ifx\csname l@#1\endcsname\relax
\typeout{** WARNING: IEEEtran.bst: No hyphenation pattern has been}%
\typeout{** loaded for the language `#1'. Using the pattern for}%
\typeout{** the default language instead.}%
\else
\language=\csname l@#1\endcsname
\fi
#2}}
\providecommand{\BIBdecl}{\relax}
\BIBdecl

\bibitem{junnan_pan_collision_2023}
\BIBentryALTinterwordspacing
J.~Pan, P.~Sotiriadis, and F.~Englberger, ``Collision {Avoidance} of {Autonomous} {Driving} at {Low} {Speed} in the {Near} {Field} of {Vehicle},'' \emph{International Journal of Emerging Engineering and Technology}, vol.~2, no.~1, pp. 57--62, Jan. 2023. [Online]. Available: \url{https://pjosr.com/index.php/ijeet/article/view/916}
\BIBentrySTDinterwordspacing

\bibitem{pan_independent_2024}
\BIBentryALTinterwordspacing
------, ``Independent {Near}-{Field} {Monitoring}: {A} {Novel} {Approach} to {Improve} {Active} {Safety} in {Autonomous} {Vehicles},'' in \emph{2024 {IEEE} {Intelligent} {Vehicles} {Symposium} ({IV})}, Jun. 2024, pp. 730--736, iSSN: 2642-7214. [Online]. Available: \url{https://ieeexplore.ieee.org/document/10588775}
\BIBentrySTDinterwordspacing

\bibitem{alsuwian_autonomous_2022}
T.~Alsuwian, R.~B. Saeed, and A.~A. Amin, ``\BIBforeignlanguage{en-US}{Autonomous {Vehicle} with {Emergency} {Braking} {Algorithm} {Based} on {Multi}-{Sensor} {Fusion} and {Super} {Twisting} {Speed} {Controller}},'' \emph{\BIBforeignlanguage{en-US}{Applied Sciences}}, vol.~12, no.~17, p. 8458, Jan. 2022.

\bibitem{yang_systematic_2022}
\BIBentryALTinterwordspacing
L.~Yang, Y.~Yang, G.~Wu, X.~Zhao, S.~Fang, X.~Liao, R.~Wang, and M.~Zhang, ``A {Systematic} {Review} of {Autonomous} {Emergency} {Braking} {System}: {Impact} {Factor}, {Technology}, and {Performance} {Evaluation},'' \emph{Journal of Advanced Transportation}, vol. 2022, pp. 1--13, Jan. 2022, publisher: Hindawi. [Online]. Available: \url{https://www.hindawi.com/journals/jat/2022/1188089/}
\BIBentrySTDinterwordspacing

\bibitem{reich_low_2024}
\BIBentryALTinterwordspacing
A.~Reich and M.~Maehlisch, ``\BIBforeignlanguage{en-US}{Low {Latency} {Instance} {Segmentation} by {Continuous} {Clustering} for {LiDAR} {Sensors}},'' in \emph{\BIBforeignlanguage{en-US}{2024 {IEEE} {Intelligent} {Vehicles} {Symposium} ({IV})}}, Jun. 2024, pp. 1871--1877, iSSN: 2642-7214. [Online]. Available: \url{https://ieeexplore.ieee.org/document/10588831}
\BIBentrySTDinterwordspacing

\bibitem{zhang_vehicle_2020}
\BIBentryALTinterwordspacing
J.~Zhang, W.~Xiao, B.~Coifman, and J.~P. Mills, ``Vehicle {Tracking} and {Speed} {Estimation} {From} {Roadside} {Lidar},'' \emph{IEEE Journal of Selected Topics in Applied Earth Observations and Remote Sensing}, vol.~13, pp. 5597--5608, 2020. [Online]. Available: \url{https://ieeexplore.ieee.org/document/9200682}
\BIBentrySTDinterwordspacing

\bibitem{pek_fail-safe_2021}
C.~Pek and M.~Althoff, ``Fail-{Safe} {Motion} {Planning} for {Online} {Verification} of {Autonomous} {Vehicles} {Using} {Convex} {Optimization},'' \emph{IEEE Transactions on Robotics}, vol.~37, no.~3, pp. 798--814, Jan. 2021.

\bibitem{jacumet_analytical_2023}
\BIBentryALTinterwordspacing
R.~Jacumet, C.~Rathgeber, and V.~Nenchev, ``Analytical {Safety} {Bounds} for {Trajectory} {Following} {Controllers} in {Autonomous} {Vehicles},'' in \emph{2023 9th {International} {Conference} on {Control}, {Decision} and {Information} {Technologies} ({CoDIT})}, Jul. 2023, pp. 730--735, iSSN: 2576-3555. [Online]. Available: \url{https://ieeexplore.ieee.org/document/10284507/?arnumber=10284507}
\BIBentrySTDinterwordspacing

\bibitem{xausa_verification_2015}
\BIBentryALTinterwordspacing
I.~Xausa, ``Verification of {Collision} {Avoidance} {Systems} using {Optimal} {Control} and {Sensitivity} {Analysis},'' Ph.D. dissertation, university of the Bundeswehr munich, Neubiberg, 2015. [Online]. Available: \url{https://eref.uni-bayreuth.de/id/eprint/31705/}
\BIBentrySTDinterwordspacing

\bibitem{yatbaz_integrity_2024}
\BIBentryALTinterwordspacing
H.~Y. Yatbaz, M.~Dianati, K.~Koufos, and R.~Woodman, ``Integrity {Monitoring} of {3D} {Object} {Detection} in {Automated} {Driving} {Systems} using {Raw} {Activation} {Patterns} and {Spatial} {Filtering},'' in \emph{2024 {IEEE} 27th {International} {Conference} on {Intelligent} {Transportation} {Systems} ({ITSC})}, Sep. 2024, pp. 161--167, iSSN: 2153-0017. [Online]. Available: \url{https://ieeexplore.ieee.org/document/10920125/}
\BIBentrySTDinterwordspacing

\bibitem{poul_greibe_braking_2007}
\BIBentryALTinterwordspacing
P.~Greibe, ``Braking distance, friction and behaviour,'' Trafitec, Tech. Rep., Jan. 2007. [Online]. Available: \url{https://www.trafitec.dk/sites/default/files/publications/braking%20distance%20-%20friction%20and%20driver%20behaviour.pdf}
\BIBentrySTDinterwordspacing

\bibitem{wang_tracking-by-detection_2022}
Y.~Wang, Z.~Wang, Y.~Huang, X.~Cui, and C.~Zheng, ``\BIBforeignlanguage{en}{A {Tracking}-{By}-{Detection} {Based} {3D} {Multiple} {Object} {Tracking} for {Autonomous} {Driving}},'' in \emph{\BIBforeignlanguage{en}{Proceedings of 2021 {International} {Conference} on {Autonomous} {Unmanned} {Systems} ({ICAUS} 2021)}}, M.~Wu, Y.~Niu, M.~Gu, and J.~Cheng, Eds.\hskip 1em plus 0.5em minus 0.4em\relax Singapore: Springer, 2022, pp. 3414--3423.

\bibitem{nenchev_compositional_2025}
\BIBentryALTinterwordspacing
V.~Nenchev, C.~Imrie, S.~Gerasimou, and R.~Calinescu, ``Compositional code-level safety verification for automated driving controllers,'' \emph{Journal of Systems and Software}, vol. 230, p. 112499, Dec. 2025. [Online]. Available: \url{https://www.sciencedirect.com/science/article/pii/S0164121225001670}
\BIBentrySTDinterwordspacing

\end{thebibliography}





\end{document}